# Data-driven study of the enthalpy of mixing in the liquid phase

Guillaume Deffrennes[a,*], Bengt Hallstedt[b], Taichi Abe[c], Quentin Bizot[a], Evelyne Fischer[a], Jean-Marc Joubert[d], Kei Terayama[e,f], and Ryo Tamura[g,h,*]

[a] Univ. Grenoble Alpes, CNRS, Grenoble INP, SIMaP, F-38000 Grenoble, France

[b] Institute for Materials Applications in Mechanical Engineering (IWM), RWTH Aachen University, Augustinerbach 4, 52062, Aachen, Germany

[c] Research Center for Structural Materials, National Institute for Materials Science, 1-2-1 Sengen, Tsukuba, Ibaraki 305-0047, Japan

[d] Univ. Paris Est Creteil, CNRS, ICMPE, UMR 7182, 2 rue Henri Dunant, 94320 Thiais, France

[e] Graduate School of Medical Life Science, Yokohama City University, 1-7-29, Suehiro-cho, Tsurumi-ku, Kanagawa 230-0045, Japan

[f] MDX Research Center for Element Strategy, Tokyo Institute of Technology, 4259 Nagatsuta-cho, Midori-ku, Yokohama, Kanagawa, 226-8501, Japan

[g] Center for Basic Research on Materials, National Institute for Materials Science, 1-1 Namiki, Tsukuba, Ibaraki, 305-0044, Japan

[h] Graduate School of Frontier Sciences, The University of Tokyo, 5-1-5 Kashiwa-no-ha, Kashiwa, Chiba 277-8561, Japan

[*] Corresponding authors at: Guillaume.DEFFRENNES@cnrs.fr (G. Deffrennes), TAMURA.Ryo@nims.go.jp (R. Tamura).

# Abstract

The enthalpy of mixing in the liquid phase is a thermodynamic property reflecting interactions between elements that is key to predict phase transformations. Widely used models exist to predict it, but they have never been systematically evaluated. To address this, we collect a large amount of enthalpy of mixing data in binary liquids from a review of about 1000 thermodynamic evaluations. This allows us to clarify the prediction accuracy of Miedema's model which is state-of-the-art. We show that more accurate predictions can be obtained from a machine learning model based on LightGBM, and we provide them in 2415 binary systems. The data we collect also allows us to evaluate another empirical model to predict the excess heat capacity that we apply to 2211 binary liquids. We then extend the data collection to ternary metallic liquids and find that, when mixing is exothermic, extrapolations from the binary systems by Muggianu's model systematically lead to slight overestimations of roughly 10% close to the equimolar composition. Therefore, our LightGBM model can provide reasonable estimates for ternary alloys and, by extension, for multicomponent alloys. Our findings extracted from rich datasets can be used to feed thermodynamic, empirical and machine learning models for material development.

# Introduction

The enthalpy of mixing provides basic information on the nature, attractive or repulsive, of the interactions between elements in solutions. Therefore, it provides important information for understanding chemical-related issues in materials, such as the glass-forming ability of metallic glasses [1]. Moreover, it is an essential property for predicting phase transformations in materials. Generally, deep eutectics, solid solutions and compounds are found in systems where mixing in the liquid phase is exothermic, e.g., Au-Sn [2], whereas miscibility gaps form in those where mixing is endothermic, e.g., Ag-Ni [3]. The enthalpy of mixing in the liquid phase is used to determine the stability of liquids by thermodynamic calculations using the Calphad technique [4], which is important for the processing and design of a large variety of alloys [5]. In addition, several empirical rules for alloy design are based on this property [6–8]. Besides, the thermodynamic properties of mixing in the liquid phase are important features for machine learning (ML) models to predict phase equilibria such as binary liquidus [5], ternary isothermal sections [9], or phase formation in high-entropy alloys [10–12].

At present, the most reliable way [13] to estimate the enthalpy of mixing in the liquid phase is to use the semi-empirical model established by Miedema *et al.* in 1980 [14]. It is remarkable that no better model has been developed since then. Miedema's model is widely used in ML models for predicting phase equilibria [5,10–12], and sometimes used in Calphad models in systems where experimental data is not available [15,16]. However, since it has never been systematically evaluated, the accuracy that can be expected from it is not well known, except for binary systems of magnesium [17].

In multicomponent liquids, data are scare. Therefore, in ML [5,10–12], empirical [6–8] and high-throughput [18,19] approaches, their enthalpy of mixing is extrapolated from the binary systems. This can be done by different geometric models [20], but the most widely used is the one proposed by Muggianu *et al.* [21] because of its simplicity. Again, the accuracy of this extrapolation has never been systematically studied.

In recent studies [18,19], the enthalpy of mixing in solid solutions was studied through high-throughput density functional theory calculations. It was approximated in binary systems from a single calculation performed at the equimolar composition. The underlying assumption is that the enthalpy of mixing is symmetrical with respect to composition and reaches its extremum at the equimolar composition. This introduces an error of unknown magnitude.

High-throughput *ab initio* approaches are currently not available to study the enthalpy of mixing of liquids. ML approaches represent an unexplored alternative. For solid solutions, a model was trained on data from density functional theory calculations to predict the enthalpy of mixing in the body-centered cubic phase for Fe-Cr-based alloys [22]. A neural network was developed to reproduce the enthalpy of mixing in binary and ternary face-centered cubic solutions obtained from a Calphad commercial database for high entropy alloys [23].

In this work, the enthalpy of mixing in the liquid phase is systematically studied using large datasets collected from a review of about 1000 thermodynamic Calphad assessments. First, we evaluate Miedema's model, and we develop a machine learning model that performs better. Second, we evaluate the performance of Witusiewicz and Sommer's empirical model [24] for predicting the temperature dependance of the enthalpy of mixing, i.e., the excess heat capacity. Third, we evaluate how accurately the enthalpy of mixing of ternary metallic liquids can be extrapolated from their binary subsystems by Muggianu's model.

## Methods

**Datasets**

About 1000 Calphad-type thermodynamic evaluations of binary systems were reviewed, and enthalpy of mixing data were collected in 375 binary systems in composition domains where the models are supported by experimental measurements with steps of 1 at%. 94% of these data are supported by direct calorimetric measurements. The remaining 6% are supported by activity or chemical potential data at different temperatures separated by at least 350 K (the contribution from the configurational entropy is then more than 2 kJ/mol), or data on both the liquidus and the thermodynamic properties of the solid phases involved. Calphad model parameters were retrieved from various sources: mainly NIMS CPDDB [25] and compilations by the authors, but also TDBDB [26], NUCLEA [27], and publications. When the liquid phase was described using a substitutional solution model, the enthalpy of mixing was calculated directly from the Redlich-Kister coefficients. For cases involving an associate or ionic model, the Pandat software [28] was used. When the enthalpy of mixing is temperature dependent, it is calculated at the composition-weighted average melting point of the elements.

Excess heat capacity data were obtained from measurements of the enthalpy of mixing at different temperatures that were digitized from figures found in publications. Data were collected in 43 binary systems for which it was found that a clear and consistent trend emerged from the experimental data. They were collected at the composition where the enthalpy of mixing is at its maximum in absolute value, or at the closest composition where experimental data are available.

Data on ternary liquids were collected from publications in which a substitutional solution model is proposed based on experimental data covering a large part of a ternary system. We cannot cite them all here, but several studies are from Jendrzejczyk-Handzlik *et al.*, such as Ref. [29], and Dębski *et al.*, such as Ref. [30].

**Miedema's model and descriptors**

Miedema's theory is presented in detail elsewhere [14,31,32]. The equations of the model and the origin of the parameters used in this work are given in the appendix.

58 descriptors are obtained from composition-weighted mean and average deviation of the properties of the pure elements as already detailed in Ref. [5]. They are listed in Table 2 of that publication. The melting and thermodynamic properties of the pure elements are obtained from the SGTE unary database [33], and other properties are calculated from magpie [34] using matminer [35].

**Machine learning methods**

Three algorithms are used: LightGBM [36], a gradient boosting decision tree, Gaussian process (GP) implemented using PHYSBO [37], and artificial neural network (ANN) with two to three hidden layers comprising 10 to 1200 nodes implemented using TensorFlow [38]. For LightGBM and ANN, the target variable is the error of Miedema's model from which the enthalpy of mixing is recalculated, and for GP it is the enthalpy of mixing directly and Miedema's model is used as a descriptor, since this leads to better results.

First, feature selection is performed based on a 5-fold group cross-validation (CV), with data for a binary system forming a group that cannot be split between the training and validation sets. For LightGBM and ANN, a preliminary hyperparameter tuning is performed using Optuna [39]. For LightGBM, recursive feature elimination is performed: starting from all 58 descriptors, the least important feature is dropped. This is repeated until only two features remain. For ANN and GP, sequential feature selection is performed: starting from a single-descriptor model, descriptors are tested one by one, and the one that leads to the model with the smaller RMSE is selected. This is repeated until a model based on all 58 or 59 descriptors is obtained for ANN or GP, respectively. This method requires a relatively high computation time, which is why feature selection was carried out as a first step on the whole dataset. The set of features that minimizes the RMSE is selected.

Second, a nested CV is performed as schematized in Fig. S3 of Ref. [5]. The advantage of this approach is that model performance is evaluated over the entire dataset. For LightGBM and ANN, a 5-fold group CV is performed in the inner loop to determine the best hyperparameters using Optuna [39]. A 12-fold group CV is performed in the outer loop to evaluate model performance.

**Redlich-Kister substitutional solution model**

For each binary system, the enthalpy of mixing is calculated at every at% from the value predicted by Miedema's model and the error on this value predicted by machine learning. Then, it is fitted by a Redlich-Kister polynomial:

$$H_{mix}^{liq} = x_A x_B \sum_v {}^v a_{A,B} (x_A - x_B)^v \qquad (1)$$

with $H_{mix}^{liq}$ the enthalpy of mixing in the A-B binary liquid phase, $x_A$ and $x_B$ the atomic fraction of A and B, and ${}^v a_{A,B}$ the parameter of order $v$. Starting from the order 0, additional parameters are iteratively added up to the order 3 if the RMSE between the predicted and fitted enthalpy of mixing is higher than 0.5 kJ/mol. In systems where the maximum absolute value of the enthalpy of mixing is predicted to be less than 10 kJ/mol, no more than two parameters of order 0 and 1 are used.

In a multicomponent liquid composed of $n$ elements, the binary contributions to the enthalpy of mixing can be calculated using Muggianu's model [21] by summing Eq. (1) over all binaries without weighting compositions:

$$^{bin}H_{mix}^{liq} = \sum_{i=1}^{n-1} \sum_{j=i+1}^{n} x_i x_j \sum_v {}^v a_{i,j}^{\varphi} (x_i - x_j)^v \qquad (2)$$

The excess heat capacity predicted by Witusiewicz and Sommer's model [24] using our LightGBM/RK model, melting points from the SGTE unary database [33], a temperature set to the composition-weighted average melting point of the elements, and boiling points from magpie [34] as inputs is fitted and can be extrapolated using the same formalism as Eqs. (1-2).

# Results

**Prediction of the enthalpy of mixing in binary liquid phases**

Reliable enthalpy of mixing data were collected in 375 binary liquids after a review of about 1000 systems. Data coverage is shown per binary system in Fig. 1, and per element in Fig. S1. In general, our dataset covers many binary systems of base alloy elements such as Al or Mg, low melting point metals such as Ga or Sn, and between elements of groups 10 to 14. In contrast, our dataset contains little information, if any, on binaries of refractory elements such as C or W, and of elements of groups 3 to 6 besides with elements of groups 8 to 14.

First, the performance of the Miedema model was evaluated on our dataset. The relation between the observed and predicted values is shown in Fig. 2a, and a mean absolute error (MAE) of 4.2 kJ/mol is obtained. In systems where the enthalpy of mixing is exothermic, it generally gives values that are not sufficiently negative.

Next, we developed an ML model based on LightGBM to predict the enthalpy of mixing. It is based on 30 descriptors selected from the 58 descriptors proposed in our previous study [5] to obtain a better accuracy (Fig. S3). Its performance is evaluated on the entire dataset using a nested cross-validation approach, and the results are shown in Fig. 2b. The values predicted by the LightGBM model are slightly noisy (Fig. S2) and fitting by a Redlich-Kister (RK) polynomial smooth them out without any significant impact on performance. The combination of the LightGBM and RK models is referred to as LightGBM/RK. The MAE of this model is of 3.0 kJ/mol in systems for which no data is included in the training set, which is better than that obtained with Miedema's model, while training data are reproduced closely with a training MAE of 0.4 kJ/mol for the final LightGBM/RK model. The mean absolute percentage error (MAPE) is an intuitive metric, but it is only meaningful for large values that are not close to zero. It is evaluated on data greater than 10 kJ/mol in absolute value from 199 binary systems. An error of 22% is obtained with our LightGBM/RK model in systems where it has seen no data, compared with 34% for Miedema's model. The LightGBM/RK model performs slightly better than the artificial neural network (ANN) and gaussian process (GP) models (Table 1). The MAE of the LightGBM/RK model is shown per element in Fig. S1. Predictions are relatively inaccurate for group 16. However, the LightGBM/RK model still tends to outperform Miedema's model: its MAE over the 21 binaries of Te in the dataset is of 6.4 kJ/mol compared to 10.4 kJ/mol for Miedema's model. Excluding group 16 elements, the MAE of the LightGBM/RK model decreases to 2.6 kJ/mol.

There are 120 binary systems in our dataset for which the extremum in the enthalpy of mixing is well-defined in the sense that it is greater than 5 kJ/mol in absolute value and that experimental data are available over the whole compositional range. In these systems, the composition of the extremum differs from the equimolar composition by 8.6 at% in average, and by up to 30 at%. Miedema's model generally fails to account for an asymmetry with respect to composition (Fig. 3), with an MAE on the composition of the extremum of 7.4 at% on these 120 systems. While our ML model show some potential to correct this (Fig. 3), this MAE is only reduced to 6.7 at% on these systems when they are not included in the training set.

Table 1: Performance metrics obtained on data from 375 binary systems for different models for predicting the enthalpy of mixing in binary liquid. RMSE stands for root mean squared error.

| Model | MAE (kJ/mol) | RMSE (kJ/mol) |
| --- | --- | --- |
| Miedema's model | 4.2 | 7.5 |
| LightGBM/RK (test) | 3.0 | 4.9 |
| ANN (test) | 3.2 | 5.2 |
| GP (test) | 3.3 | 5.1 |

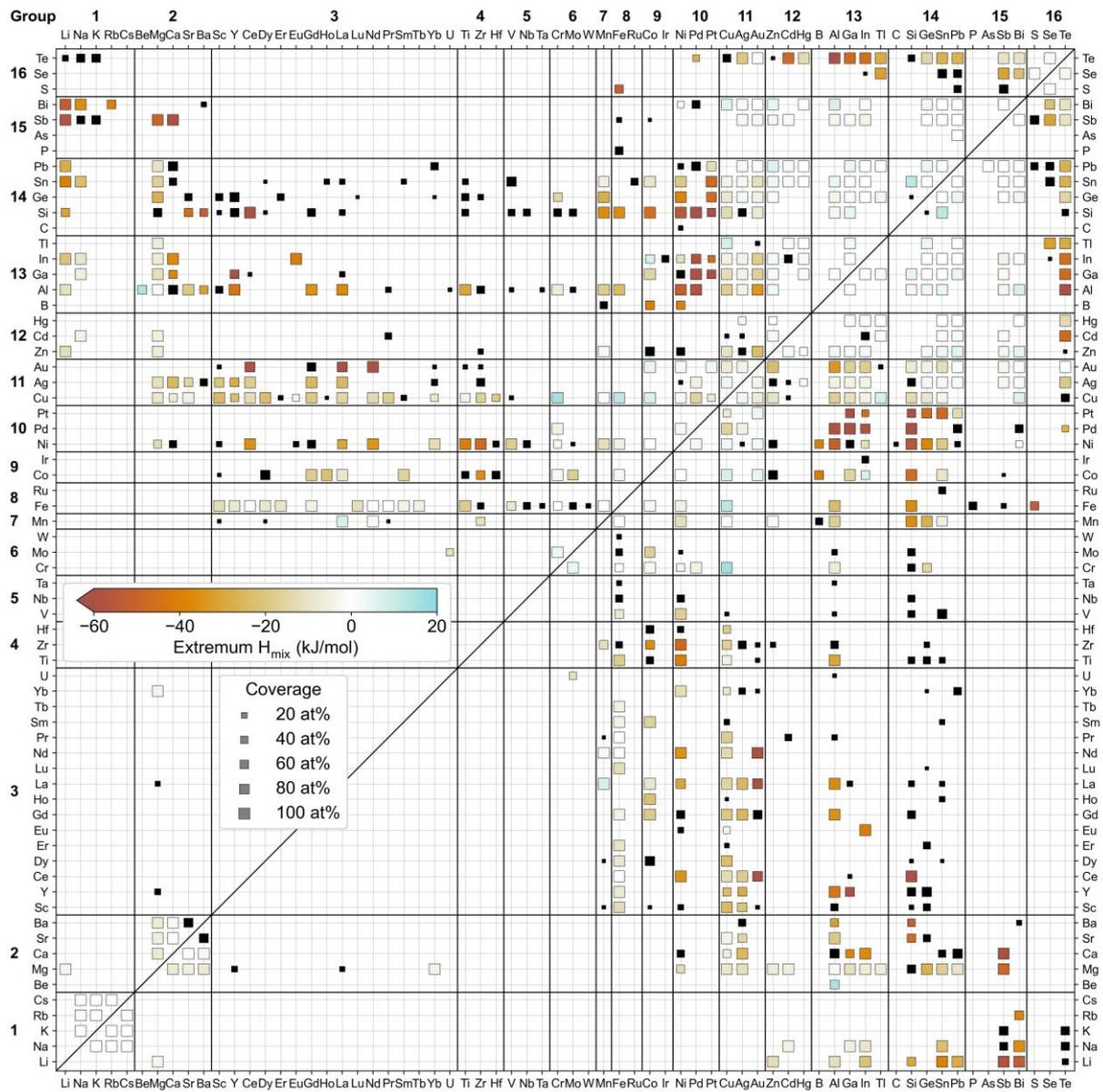

Fig. 1: Overview of our dataset on the enthalpy of mixing in binary liquid phases. For each of the 375 binary systems included, symbol size corresponds to the span of compositions over which reliable data have been found, and symbol color to the extremum value of the enthalpy of mixing. Symbols are black when the extremum is not well-defined.

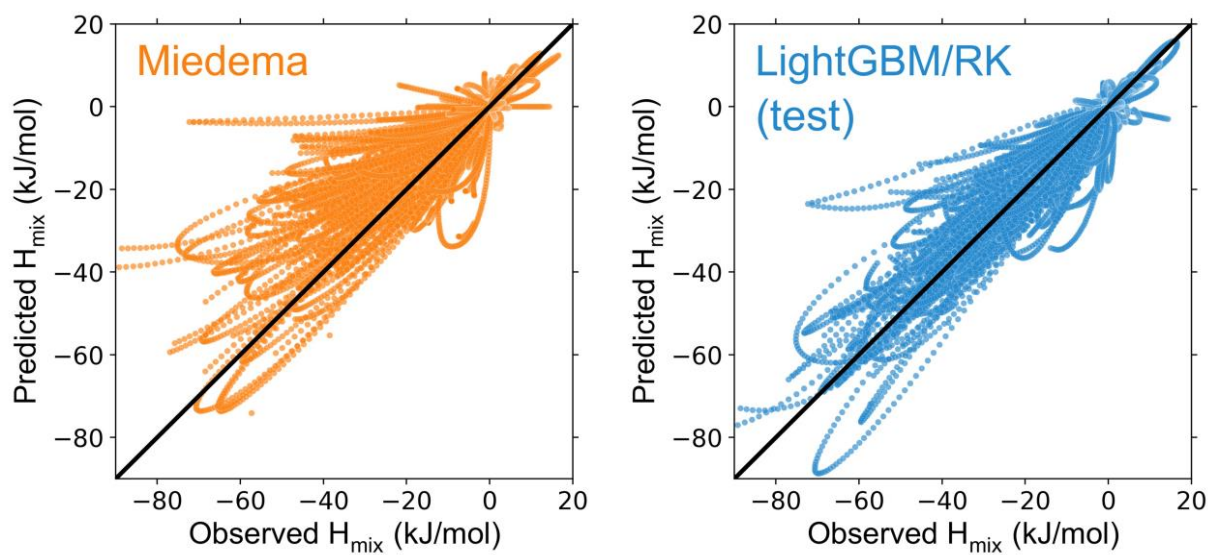

Fig. 2: Scatter plots showing the performance on data from 375 binaries of Miedema's model [14] and of our ML model (in test) after fitting by a Redlich-Kister substitutional solution model. The diagonal line represents perfect agreement.

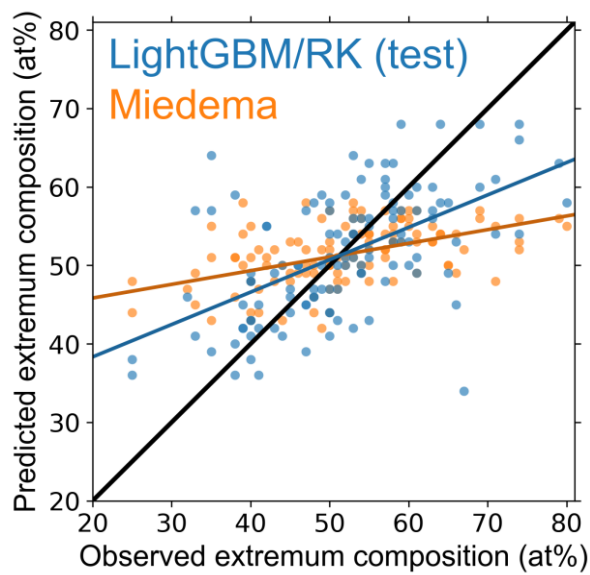

Fig. 3: Performance for predicting the composition of the extremum of Miedema's model and of our ML model (in test) on data from 120 binaries for which the extremum is well-defined. Linear regression lines are plotted to guide the eye.

The enthalpy of mixing is then predicted in all the 2415 binary systems generated by 70 elements using our LightGBM/RK model. The extremum value obtained in each system is shown in Fig. 3. The enthalpy of mixing is predicted from the LightGBM/RK model to be most exothermic in binary systems between an element of groups 2 to 4 with an element of groups 9 to 16 (Fig. 4). For instance, extremum values of -150 kJ/mol are predicted in carbon - rare earths systems. It is most endothermic in binary systems between an alkali metal and an element of groups 3 to 8. Mg and Be tend to behave more like Au and Zn, which are their neighbors in an ordering by Mendeleev number, than other alkaline-earth metals. In binary systems between two elements of the same group, the enthalpy of mixing tends to be small and no larger than 30 kJ/mol in absolute value.

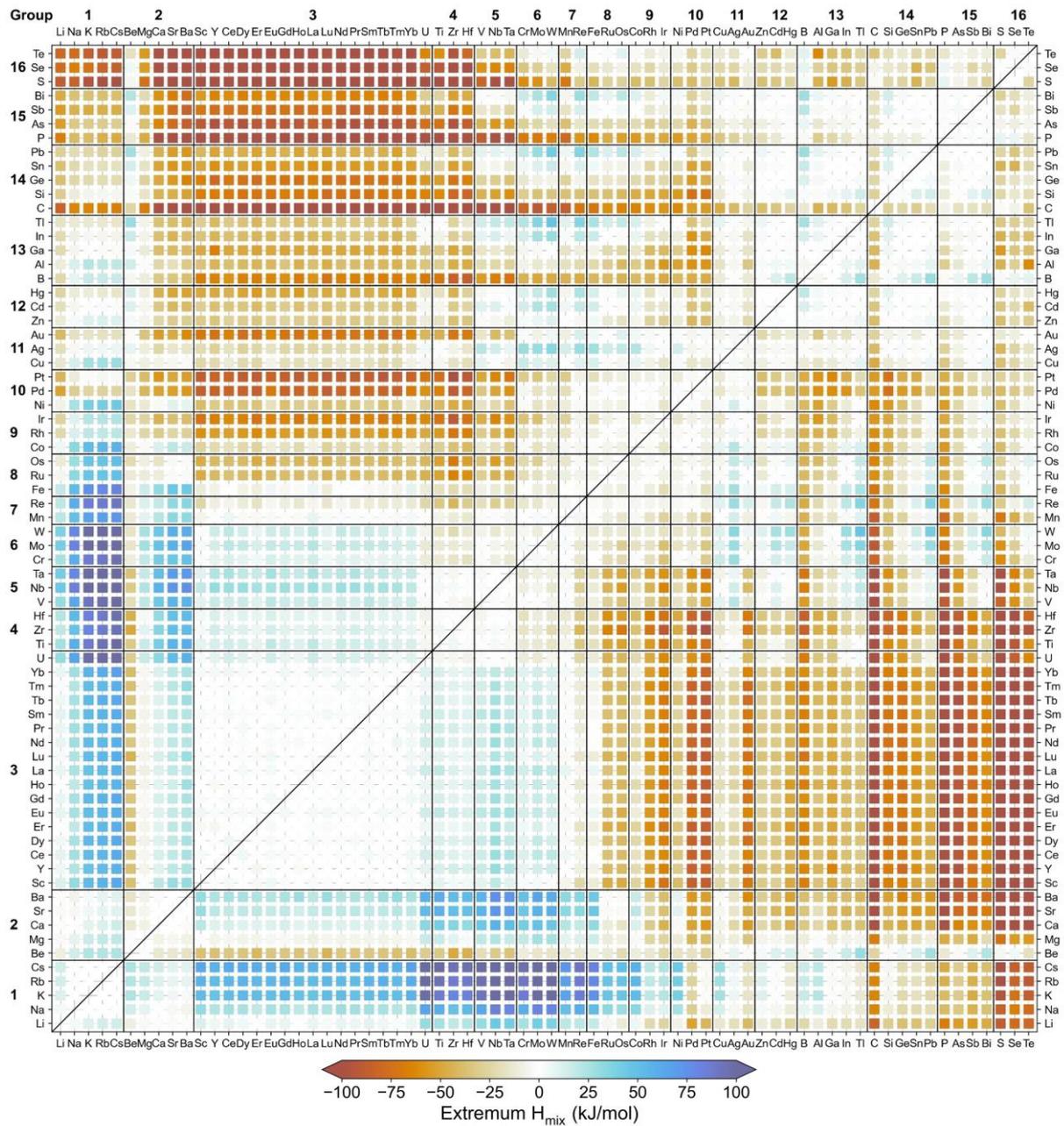

Fig. 4: Extremum value of the enthalpy of mixing in the liquid phase predicted using the LightGBM/RK model over all the 2415 binary systems generated by 70 elements.

**Temperature dependence of the enthalpy of mixing in the liquid phase**

The excess heat capacity is the temperature derivative of the enthalpy of mixing. It can be predicted from Witusiewicz and Sommer's empirical model [24] whose MAE on experimental data from 77 binary systems is of 3.7 J/K/mol (Fig. 5a). It takes the enthalpy of mixing as input, and using our LightGBM/RK model predictions can be made with greater accuracy than using Miedema's model, especially in systems included in its training set (Fig. 5b). The results obtained this way in 2016 binary systems suggests that the excess heat capacity and the enthalpy of mixing are strongly correlated (Fig. 5c). As discussed in Ref. [40], they are in most cases of opposite signs, which means that temperature brings liquids closer to ideality. However, experimental evidence is lacking in systems where mixing is endothermic, such as Ag-Pb where measurements suggest an exception to this rule [41]. In 90% of the binary systems, the extremum value of the enthalpy of mixing is predicted to decrease by 3.5 to 9.7% in absolute value over 100 K (Fig. 5c).

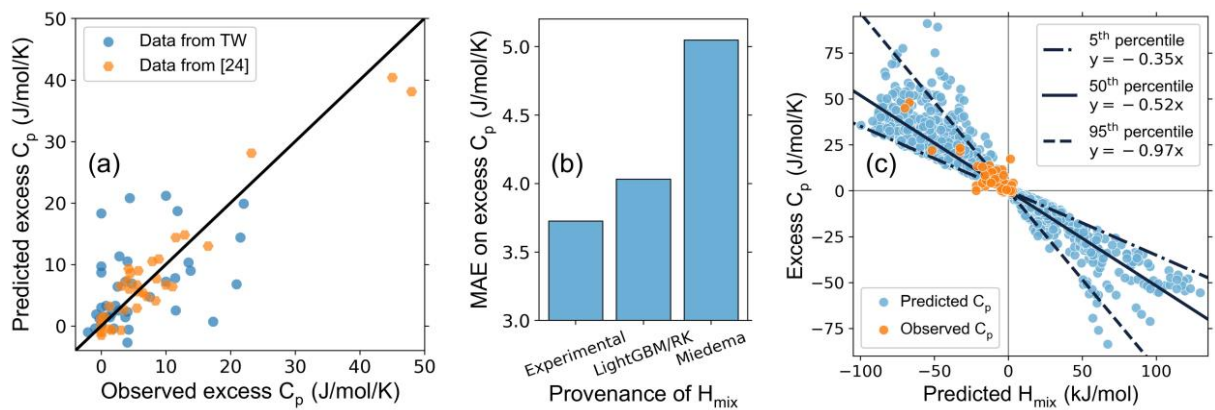

Fig. 5: Performances over data on 77 binary liquids of Witusiewicz and Sommer's model [24] using (a) experimental and (b) experimental or predicted values for the input enthalpy of mixing. (c) Relationship between the excess heat capacity and the enthalpy of mixing over 2016 binary liquids generated by 64 elements of groups 1 to 15 at the composition where the enthalpy of mixing reaches its extremum. Excess heat capacity predictions for As, C and P are abnormal due to their boiling points being lower than their melting points and are not plotted. Percentile regression lines are obtained from absolute values.

**Extrapolation from the binary systems in ternary metallic liquids**

In multicomponent liquids, the main contribution to the enthalpy of mixing comes from the binary contributions. The extent of higher-order contributions has never been studied systematically. To address this, we collected experimental data on 52 ternary metallic liquids close to the equimolar composition. We find that the enthalpy of mixing extrapolated from the binary systems by Muggianu's model [21] is reasonable, with an MAE of 1.6 kJ/mol. In liquids where mixing is exothermic, the extrapolations are systematically too negative (Fig. 6). The overestimation is of the order of 10% (Fig. S4). In liquids where mixing is endothermic, more data is necessary to draw conclusions.

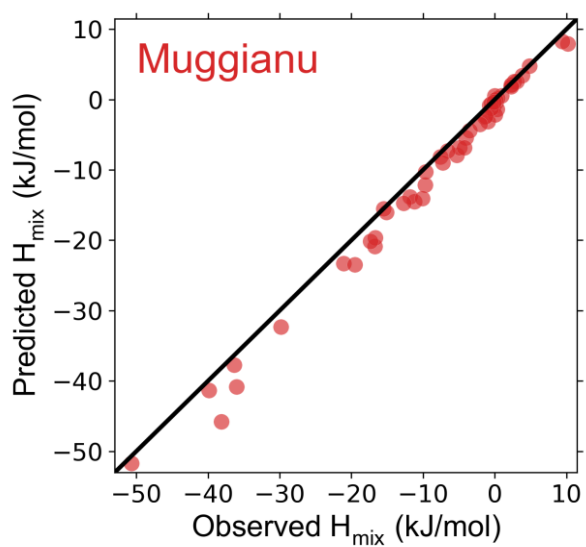

Fig. 6: Scatter plot showing the enthalpy of mixing extrapolated from the binary systems by Muggianu's model [21] against experimental data on 52 ternary near-equimolar metallic liquids.

## Discussion

We developed an ML model to predict the enthalpy of mixing in binary liquid phases. It is significantly more reliable than Miedema's model [14]. The prediction results obtained in all the 2415 binary systems generated by 70 elements are fitted by a substitutional solution model whose parameters are given in the supplementary materials. From these parameters, the main contribution to the enthalpy of mixing can be easily calculated in multicomponent liquids using Eq. 2 or the open-access notebook linked in the Data availability section. In ML models for alloys, the enthalpy of mixing in the liquid phase is an important descriptor not only for predicting phase equilibria [5,9–12], but also properties of the alloys such as the Young's modulus [42] or the Vickers hardness [43]. Our ML model is used as an input to Witusiewicz and Sommer's model [24] to predict the excess heat capacity in 2016 binary liquids. Data on the excess heat capacity are scarce, so in thermodynamic evaluations it is common practice to set it to zero and to adjust the excess entropy to model the temperature dependence of the Gibbs energy. Instead, relying on the reasonable excess heat capacity estimates provided here should in most cases lead to a better description of the entropy and of the interactions at high temperature. Therefore, our predictions can be used as an input to thermodynamic, empirical and ML models for the development of a variety of materials.

We took a data-driven approach to evaluate different sources of error in estimation of the enthalpy of mixing in binary and multicomponent liquids: model error to predict its binary contributions, neglecting its temperature dependance, and neglecting its ternary contributions. The same type of approach should be applied to solid solutions, but similar deviations may be expected. The uncertainty from measurements or calculations is difficult to evaluate systematically and is not discussed. Our findings can be summarized as follows. First, the enthalpy of mixing predicted by Miedema's model in binary liquids where mixing is exothermic is in most cases not sufficiently negative. This is not the case with our LightGBM/RK model for which an MAE of 3.0 kJ/mol, or an MAPE of 22%, can be expected in systems not included in its training set. However, its accuracy in predicting the composition of the extremum remains limited in systems for which no data is included in the training set, with an MAE of 6.7 at% obtained over 120 binaries. Second, neglecting the temperature dependence of the enthalpy of mixing can lead to an error of roughly 5% per 100 K on the extremum value (Fig. 5c). Third, in metallic liquids where mixing is exothermic, close to ternary equimolar compositions, the enthalpy of mixing extrapolated from the binary systems by Muggianu's model is systematically overestimated, i.e., too negative, by roughly 10%. The same is observed for the faced-centered cubic phase of the Co-Cr-Fe-Mn-Ni system [44], which strengthens this observation and suggests that it may also holds for solid solutions. It is concluded that none of these errors is negligible, but when estimating the enthalpy of mixing of complex concentrated alloy, the prediction of its binary contributions remains the most important source of error.

The quality of our results depends on that of our datasets. They were obtained from Calphad assessments whose reliability was critically evaluated on a case-by-case basis. The advantages of this approach were discussed in Ref. [5]. It enabled us to collect an unprecedently large amount of data supported by direct and indirect measurements. Our dataset on the enthalpy of mixing in binary liquids is centered on metals (Fig. 1). The LightGBM/RK model performs relatively poorly on group 16 elements. On test, this can be explained by the fact they tend to form complex liquids with strong short-range ordering that are underrepresented in the dataset. Their enthalpy of mixing has a characteristic V-shape, but an enthalpy with a U-shape is predicted as in metallic liquids (Fig. S2). On training, the problem comes from the use of a substitutional solution model (Eq. 1) that is not well suited for these liquids. In immiscible systems where the enthalpy of mixing is endothermic, measurements are difficult [3], and the model is only supported by a limited number of data all below 20 kJ/mol (Fig. 2). Our data on the enthalpy of mixing in binary liquids are within or close to experimental uncertainty. However, our data on the ternary contributions to the enthalpy of mixing are less reliable because they depend on the choice made regarding binary contributions. For example, in the Ag-Ga-Sn system, the maximum ternary contributions can vary from +4.8 to +2 kJ/mol on the basis of the same experimental data but a different description of the Ag-Ga binary [45,46]. Besides, data on the excess heat capacity are also less reliable since they are derived from two experimental data at different temperatures.

## Data availability

The parameters of our LightGBM/RK model obtained in the 2415 binary systems generated by 70 elements are given at [https://drive.google.com/file/d/14RTbijm_d97V-GlPN35F4Zh4hcKYsccw](https://drive.google.com/file/d/14RTbijm_d97V-GlPN35F4Zh4hcKYsccw) and [https://drive.google.com/file/d/1GANAEu6TfcAwO-J3mEDI8URjTIbCrP3y](https://drive.google.com/file/d/1GANAEu6TfcAwO-J3mEDI8URjTIbCrP3y). The binary contributions to the enthalpy of mixing in the liquid phase can be calculated in J/mol from these using Eqs. (1-2). The parameters to calculate the excess heat capacity in J/mol/K in 2211 binary liquids generated by 67 elements based on Witusiewicz and Sommer's model [24] are also given. Descriptors for machine learning models for alloys can be easily generated from our models using the open-access notebook at [https://colab.research.google.com/drive/1XRqgvP11JuU8pxHqlpkZQChAJ5tvLdVW](https://colab.research.google.com/drive/1XRqgvP11JuU8pxHqlpkZQChAJ5tvLdVW). Our datasets on the excess heat capacity and on the enthalpy of mixing in ternary liquids will be provided as supplementary materials of the published version.

## Acknowledgements

The authors would like to thank Etsuko Ogamino, Stéphane Gossé and Alexander Pisch for contributing to discussion and data collection. The authors would also like to thank the IRSN (Institut de Radioprotection et de Sûreté Nucléaire) for access to the NUCLEA database. This research was supported by the CREST (Core Research for Evolutional Science and Technology) program of the Japan Science and Technology Agency (Grant No. JPMJCR17J2) and by the MEXT Program: Data Creation and Utilization Type Material esearch and Development Project (Grant Number JPMXP1122683430). This work has been partially supported by MIAI@Grenoble Alpes (ANR-19-P3IA-0003).

## Competing interests

The authors declare no competing interests.

# Appendix. Prediction of the enthalpy of mixing in the liquid phase using Miedema's model

The enthalpy of mixing in a binary A-B liquid is predicted by Miedema's model by following steps S1 to S5:

$$c_B^s = \frac{x_B V_B^{2/3}}{x_A V_A^{2/3} + x_B V_B^{2/3}} \tag{S1}$$

with $x_B$, $V_B$ and $c_B^s$ the molar fraction, molar volume, and surface fraction of B.

$$(V_A^{2/3})_{alloy} = V_A^{\frac{2}{3}}(1 + a_A c_B^s(\varphi_A - \varphi_B)) \tag{S2}$$

with $a_A$ a parameter for A, $\varphi_B$ the so-called work function of B, and $(V_A^{2/3})_{alloy}$ the molar volume of A accounting for changes upon alloying.

$$(c_B^s)_{alloy} = \frac{x_B (V_B^{2/3})_{alloy}}{x_A (V_A^{2/3})_{alloy} + x_B (V_B^{2/3})_{alloy}} \tag{S3}$$

with $(c_B^s)_{alloy}$ the surface fraction of B corrected as recommended in Ref. [47].

$$H_{inter}(\text{A in B}) = \frac{(V_A^{2/3})_{alloy}(-P(\varphi_A - \varphi_B)^2 + Q(n_{wsA}^{1/3} - n_{wsB}^{1/3})^2 + 0.73 R_A R_B)}{\frac{1}{2}\left(\frac{1}{n_{wsA}^{1/3}} + \frac{1}{n_{wsB}^{1/3}}\right)} \tag{S4}$$

$H_{inter}(\text{A in B})$ is the interfacial enthalpy of solving one mole of A in B obtained for the liquid phase. $P$ is a parameter equal to 14.2, 12.35 or 10.7 [32] for alloys of two transition metals, a transition metal with a non-transition metal, and two non-transition metals, respectively. Fig. 23 of Ref. [14] details which elements are considered transition metals and which are not. $n_{wsA}$ is the averaged electron density at the boundary of the Wigner-Seitz cell of A. $R$ is another parameter, and 0.73 is a reducing factor used for the liquid phase. The ratio of $Q$ over $P$ equals 9.4.

The enthalpy of mixing in the A-B liquid, $H_{mix}$, is finally obtained in step 5:

$$H_{mix} = c_A c_B^S H_{inter}(\text{A in B}) \tag{S5}$$

For elements of group 1 to 15, parameter values are taken from matminer [35] and checked against Refs. [14,31]. The $R$ parameter for Ag is corrected from 0.3 to 0.15, and the *a* parameter for Eu and Yb are corrected from 0.1 to 0.07, which is the recommended value for trivalent metals [14]. For elements of group 16, parameter values are taken from Ref. [48]. The model is not straightforward to use. We have verified for a few systems that the values obtained in this work were the same as those obtained in Ref. [47]. Significantly different results can be obtained with other calculators proposed in the literature due to errors in parameters or possibly equations.

**Supplementary Note A: Additional results on the LightGBM/RK model**

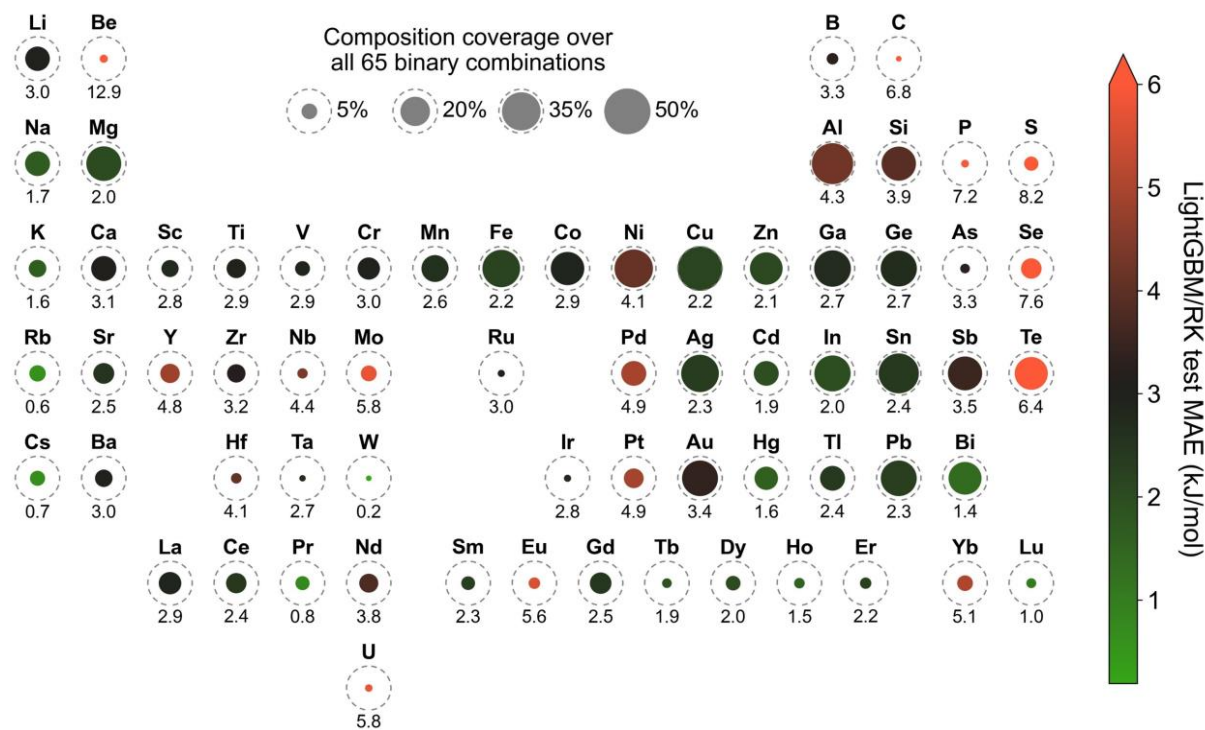

Fig. S1: Performance of the LightGBM/RK model per element.

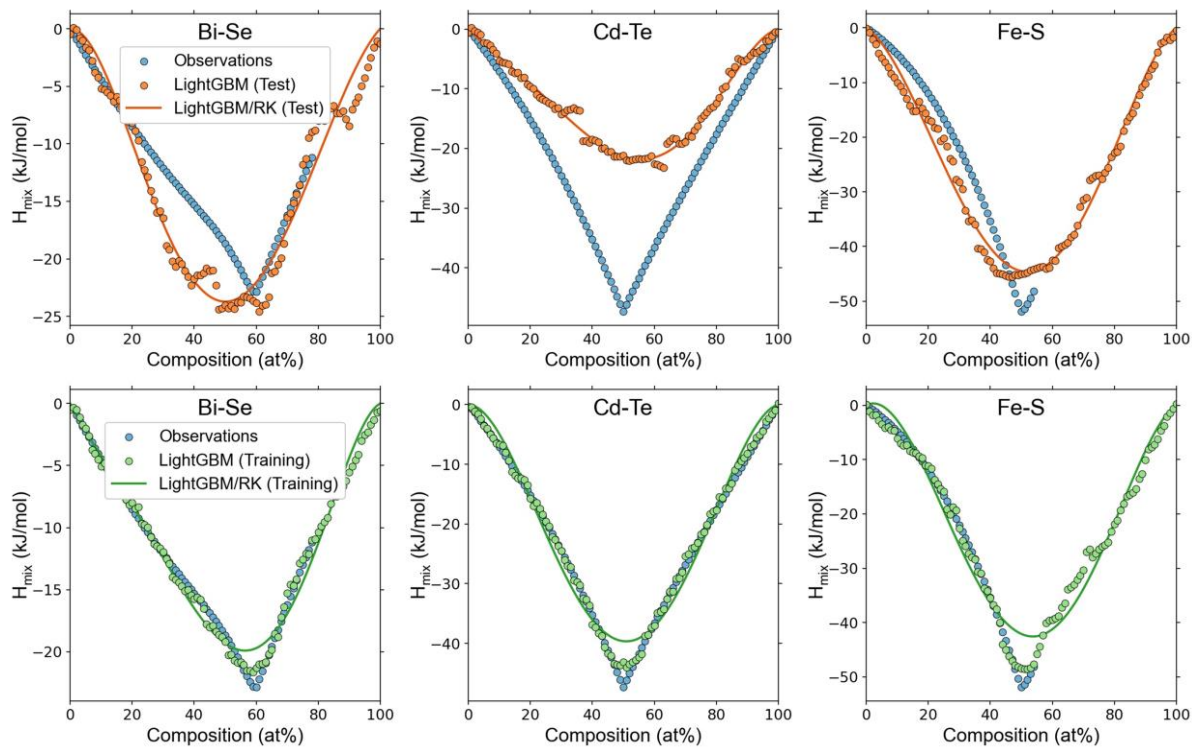

Fig. S2: Examples of training and test predictions of the enthalpy of mixing in binary liquids of group 16 elements by our LightGBM/RK model

**Supplementary Note B: Results of feature selection**

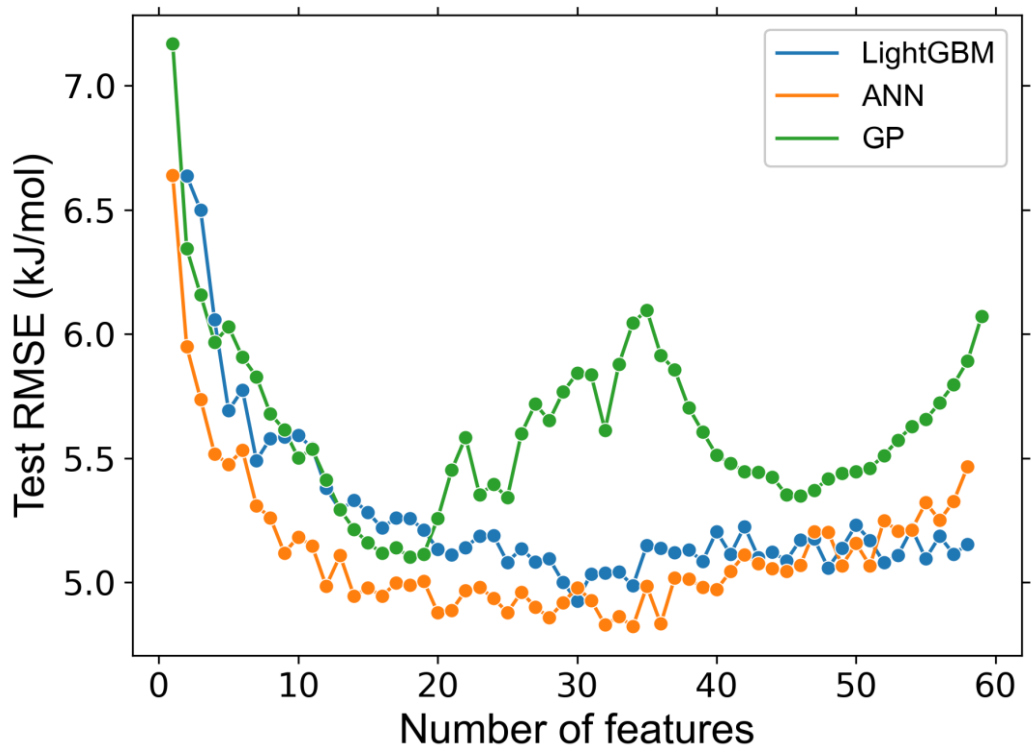

Fig. S3: Results of feature selection for different algorithms

**Supplementary Note C: Additional information on the ternary contributions to the enthalpy of mixing in metallic liquids**

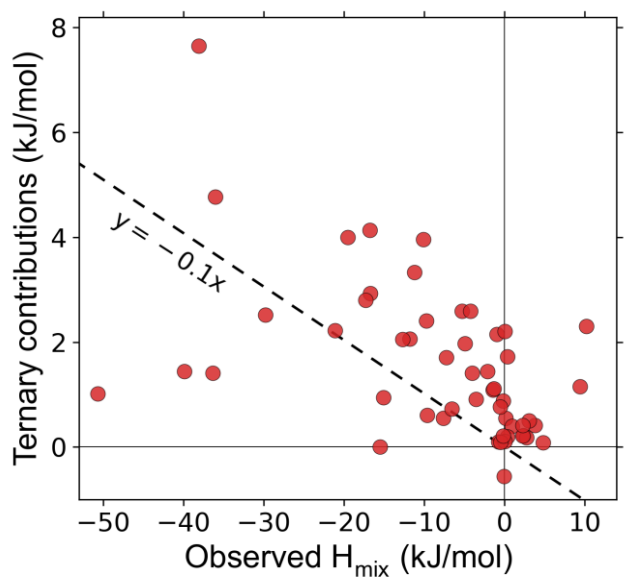

Fig. S4: Scatter plot showing experimental enthalpy of mixing data on 52 ternary near-equimolar metallic liquids against the ternary contributions modeled to correct the values extrapolated from the binary systems by Muggianu's model [21]

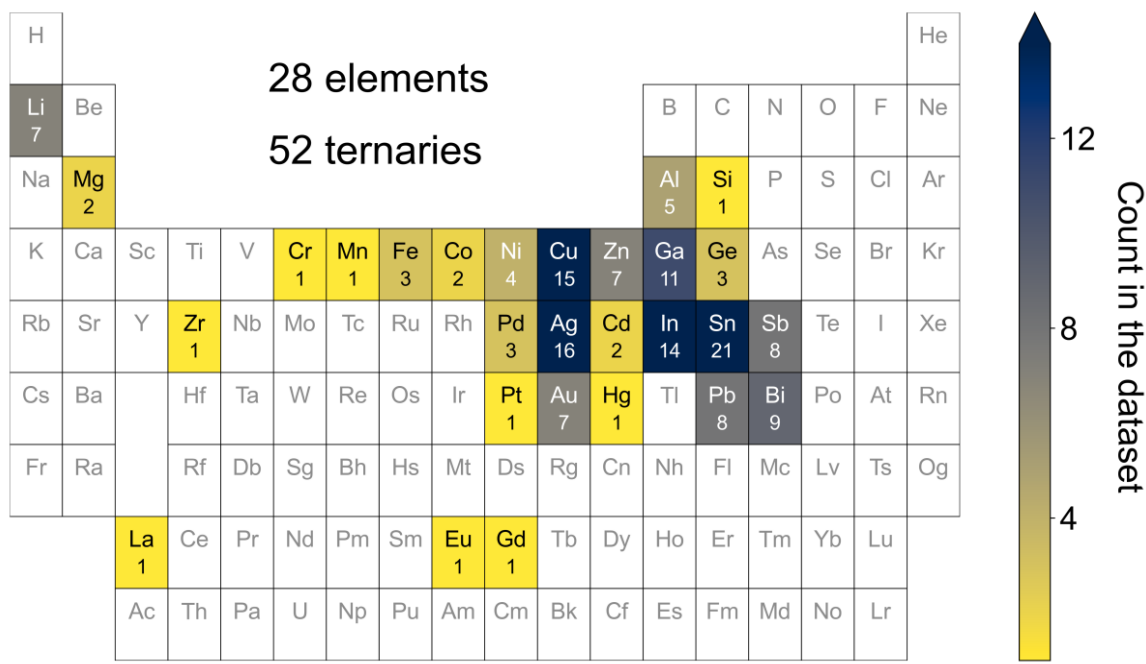

Fig. S5: Number of ternary systems included per element in our dataset on the enthalpy of mixing in ternary metallic liquids